\documentclass[conference,a4paper]{IEEEtran}
\usepackage[utf8]{inputenc}
\usepackage[T1]{fontenc}
\usepackage{url}              
\usepackage{cite}             
\usepackage[cmex10]{amsmath}
\usepackage{amssymb}
\usepackage{xifthen}
\usepackage{comment}
\usepackage{xcolor}

\usepackage{mleftright}       % fix to wrong spacing of \left-,
\mleftright                   % \middle- \right-commands 

\usepackage{graphicx}
\usepackage{caption}
\usepackage{subcaption}
\usepackage{booktabs}

\usepackage{algorithmicx}       %In original 
\usepackage{algorithm}          %had to add since not imported by algorithmicx
\usepackage{algpseudocode}

\usepackage{amsthm} %Had to use it even if the given template does not allow since the proof environment is not defined otherwise

\newtheorem{theorem}{\bfseries Theorem}%[section]
\newtheorem{lemma}{\bfseries Lemma}
\newtheorem{definition}{\bfseries Definition}
\newtheorem{corollary}{\bfseries Corollary}

\newtheorem{remark}{\bfseries Remark}

\newtheorem{example}{\bfseries Example}

\newtheorem*{assumptions*}{\bfseries Assumptions}

\usepackage{mathtools}

\DeclareMathOperator{\diag}{diag}

\DeclareMathOperator{\IID}{i.i.d}
\DeclareMathOperator{\flag}{flag}

\newcommand{\p}[1]{p_{x}}
\newcommand{\q}[1]{q_{\hat{x}}^{#1}}
\newcommand{\Q}[1]{Q_{\hat{x}|x}^{#1}}
\newcommand{\Qm}[1]{Q_{\hat{x}}^{#1}}
\renewcommand{\v}[1]{v^{#1}}
\newcommand{\tq}[1]{\tilde{q}_{\hat{x}}^{#1}}
\newcommand{\tQ}[1]{\tilde{Q}_{\hat{x}|x}^{#1}}
\newcommand{\tQm}[1]{\tilde{Q}_{\hat{x}}^{#1}}
\newcommand{\tW}{\tilde{W}}
\newcommand{\df}{\partial f}

\newcommand{\setX}{\mathcal{X}}
\newcommand{\setXh}{\mathcal{\hat{X}}}

\newcommand{\A}[2][]{
    \ifthenelse{\isempty{#1}} 
        {A^{#2}(x,\hat{x},s)} 
        {A^{#2}(x,#1,s)}
}

\newcommand{\tA}[2][]{
    \ifthenelse{\isempty{#1}} 
        {\tilde{A}^{#2}(x,\hat{x},s)} 
        {\tilde{A}^{#2}(x,#1,s)}
}

\newcommand{\dsum}[2]{ 
    \sum_{(#1,#2) \in \mathcal{X} \times \mathcal{\hat{X}}}
}

\newcommand{\ssum}[1]{ 
    \sum_{#1 \in \mathcal{X}}
}

\newcommand{\ssumh}[1]{ 
    \sum_{#1 \in \mathcal{\hat{X}}}
}

\newcommand{\E}[2][]{ 
    \mathbb{E}_{#1} \left[ #2 \right]
}

\newcommand{\cx}[1][]{
    \ifthenelse{\isempty{#1}} 
        {c(\hat{x})} 
        {c(#1)}
}

\begin{document}

\title{Computation of Rate-Distortion-Perception Function under $f$-Divergence Perception Constraints} 

\author{%
   \IEEEauthorblockN{Giuseppe Serra,
                    Photios A. Stavrou,
                    Marios Kountouris}
                    %\IEEEauthorrefmark{4},
                    % and Harry Potter\IEEEauthorrefmark{1}}
  \IEEEauthorblockA{Communication Systems Department, EURECOM, Sophia-Antipolis, France %Hogwarts School of Witchcraft and Wizardry,
%                     1714 Hogsmeade, Scotland,
                    %\{fotios.stavrou, marios.kountouris\}@eurecom.fr
                    }
\IEEEauthorblockA{\texttt{\{giuseppe.serra,fotios.stavrou,marios.kountouris\}}@eurecom.fr}
}

\maketitle

\begin{abstract}
 In this paper, we study the computation of the rate-distortion-perception function (RDPF) for discrete memoryless sources subject to a single-letter average distortion constraint and a perception constraint that belongs to the family of $f$-divergences. For that, we leverage the fact that RDPF, assuming mild regularity conditions on the perception constraint, forms a convex programming problem. We first develop parametric characterizations of the optimal solution and utilize them in an alternating minimization approach for which we prove convergence guarantees. The resulting structure of the iterations of the alternating minimization approach renders the implementation of a generalized Blahut-Arimoto (BA) type of algorithm infeasible. To overcome this difficulty, we propose a relaxed formulation of the structure of the iterations in the alternating minimization approach, which allows for the implementation of an approximate iterative scheme. This approximation is shown, via the derivation of necessary and sufficient conditions, to guarantee convergence to a globally optimal solution. We also provide sufficient conditions on the distortion and the perception constraints which guarantee that our algorithm converges exponentially fast. We corroborate our theoretical results with numerical simulations, and we draw connections with existing results. 
\end{abstract}

\section{Introduction}

\par The theoretical framework of rate distortion theory subject to a single-letter average distortion constraint stems from the seminal work of Shannon in \cite{6773024,shannon:59}. Therein, Shannon described, for the first time, the fundamental trade-offs between the desired bit rate used for a compressed representation of the source messages and the associated achievable distortion criterion attained between the source message and its reconstructed representation.  The mathematical representation in rate distortion theory is manifested by the rate-distortion function (RDF). Rate distortion theory has been the foundational basis to study and develop lossy compression algorithms for various multimedia applications.
\par In recent years, it has been shown through multiple works, spanning from machine learning and computer vision to multimedia applications \cite{Minnen:2018, Khisti:2021:GaussianRDP, Chen:2021:improc_ml, theis:2016:hyperRes, shaham:2018:defImCompr, kudo:2019}, that focusing exclusively on distortion minimization does not necessarily imply a good perceptual quality of the reconstructed signal, where perceptual quality refers to the property of a sample to appear visually pleasing from a human perspective. Motivated by the need for a general characterization of RDF to enable encompassing the perceptual quality of the sample, Blau and Michaeli in \cite{BlauMichaeli_RDP} introduced a generalization of the single-letter RDF, coined rate-distortion-perception function (RDPF). The specific information theoretic characterization complements the classical single-letter distortion constraint between a source message and its reconstruction, which is inherent in RDF formulation, with a divergence constraint between the induced distributions of the source and its estimated value. The divergence constraint in RDPF is precisely used as a proxy of the human perception, measuring the degree of satisfaction in the consumption of the data from a human perspective. The divergence constraint can also be viewed as a semantic quality metric, which measures the degree of relevance of the reconstructed source from the perspective of the observer, a point first hinted at in \cite{kount2020sem}. Another relevant yet different setup is the recently introduced robust source coding framework, see, e.g., \cite{stavrou:2023} (and references therein), in which in place of the perception quality, there exists an additional distortion criterion. %One major difference between RDPF and the setup in \cite{stavrou:2023} is that in the former the characterizations are solved for various examples using separately each fidelity constraint, whereas in the latter, one can study from an optimization standpoint the joint behavior of the two distortion penalties.} 
\par Since its conception, RDPF has received substantial interest from the information theory community. Theis and Wagner in \cite{theis:2021} proved that the RDPF can be achieved via stochastic, variable-length codes by making use of a strong functional representation lemma \cite{li:2018}, whereas Chen {\it et al.} in \cite{Chen_RDPAchivability} studied the role of stochasticity in the encoder/decoder structure and proved the achievability of the RDPF by deterministic codes (except certain extreme cases) in the asymptotic regime. Wagner in \cite{wagner:2022} considered the case of perfect perceptual quality in the definition of RDPF and devised a coding theorem that allows for a specified amount of common randomness between the encoder and decoder.
\par It should be emphasized that similarly to the classical RDF, there are no known closed-form representations of the RDPF for general alphabet sources. In fact, there are only a few examples in the literature where analytical expressions are provided for the RDPF, such as the case of binary sources subject to Hamming distortion and total variation distance \cite{BlauMichaeli_RDP,Kristi_20.500.11850/535275} or the case of a Gaussian source subject to a mean-squared error distortion and a 2-Wasserstein distance \cite{Khisti:2021:GaussianRDP}. 
\par In this paper, we focus on the computation of the RDPF for discrete memoryless sources subject to a single-letter average distortion constraint and a perception constraint that belongs to the class of $f$-divergences\footnote{For a mathematical background on $f$-divergences and their properties we refer the reader to \cite{sason:2018}.}. In particular, we leverage the fact that under mild regularity conditions on the perception constraint (i.e., convexity on the second argument) the RDPF forms a convex program. This enables us to derive a parametric characterization of the optimal solution of the RDPF (Lemma \ref{lemma: Minimization_Conditions}), which is subsequently utilized to construct an alternating minimization procedure\footnote{For details on the alternating minimization procedure, see e.g., \cite{csiszar:1984,yeung:2008}.} for which we also establish convergence guarantees (Theorem \ref{th:BA Convergence}). The resulting structure  of the iterations in Theorem \ref{th:BA Convergence} prohibits the implementation of a generic Blahut-Arimoto (BA) algorithm similar to what is already known for the classical rate distortion theory for $\IID$ sources and single-letter distortions \cite{Blahut_1054855}. To overcome this technical difficulty, we introduce a new relaxed formulation of the structure of the iterations in Theorem \ref{th:BA Convergence}, which results in a variant of an approximate BA algorithm (Theorem \ref{th:BA Convergence Approximate}). Additionally, in Theorem \ref{th:BA Convergence Approximate}, we derive necessary and sufficient conditions to ensure that our approximation algorithm converges to a globally optimal solution. In light of our result in Theorem \ref{th:BA Convergence Approximate}, we also provide sufficient conditions on the structure of the distortion and the perception constraints based on which our algorithm converges exponentially fast (Theorems \ref{th: EXPConvergenceBase}, \ref{th: approximation_exponential_interval}). We corroborate our theoretical findings with numerical simulations and draw comparisons with existing results in the literature (Section \ref{sec: Numerical Examples}).

\paragraph*{\bf Notation} We denote with $\p{}(x)$ the probability distribution over the source alphabet $\mathcal{X}$ evaluated on the symbol $x$, and by $\q{}(\hat{x})$ any arbitrary marginal probability on the output alphabet $\mathcal{\hat{X}}$ evaluated on the symbol $\hat{x}$. $\Q{}(\hat{x}|x)$ denotes the entry $(\hat{x},x)$ of the transition matrix $\Q{}$ while $\Qm{}(\hat{x})$ denotes the marginal on $\mathcal{\hat{X}}$ induced by $\p{}$ and $\Q{}$, evaluated on the symbol $\hat{x}$.  We denote by $\E[]{\cdot}$ the expectation operator, and by $\E[q]{\cdot}$ the probability distribution $q$ on which the expectation operator is applied. We denote by $p_1[p_2]$ the explicit form of the functional dependence of some probability distribution $p_1$ functionally dependent on another probability distribution $p_2$. We denote by $D(\cdot||\cdot)$ any divergence measure and by $D_f(\cdot||\cdot)$, $D_{KL}(\cdot||\cdot)$, $D_{\chi^2}(\cdot||\cdot)$, $TV(\cdot||\cdot)$, the class of $f$-divergence, the Kullback–Leibler divergence, the chi-squared divergence and the total variation distance, respectively. 
We denote by $f \in C^0$, the continuous function defining the  $f$-divergence and $\df$ denotes its sub-gradient \cite[Definition 8.3]{rockafellar:1998}. For a continuous and twice differentiable function $f \in C^2$, we denote by $f''(\cdot)$ the second derivative with respect to its argument.

\section{Preliminaries on the RDPF}
In this section, we consider finite alphabets sources and deterministic encoders/decoders pairs and define the minimum achievable rates subject to average per-letter distortion and average per-letter perception constraints. Our analysis herein stems from recent results in \cite{Chen_RDPAchivability}.

\par We assume that we are given an $\IID$ sequence of $n$-length random variables ${X}^n$ that induce the probability distribution $\p{}$. The source sequence is received by an encoder $(e)$ that generates the index $e(X^n)\in{\cal Z}$, ${\cal Z}=\{1,2,\ldots,2^{nR}\}$, whereas at the decoder $(g)$, the system reconstructs an estimate of $\hat{X}^n$. Formally, the encoder and decoder are deterministic mappings with $e: \setX^n \to \mathcal{Z}$ and $g:\mathcal{Z} \to \setXh^n$, respectively.
\par  We let $d: \mathcal{X} \times \mathcal{\hat{X}} \to \mathcal{R}_0^+$ to denote a single-letter distortion function and $D: \mathcal{P} \times \mathcal{Q} \to \mathcal{R}_0^+$ to denote a divergence measure.
 Moreover, we define the fidelity criterion $\Delta$ as the expected per-symbol distortion
and the fidelity criterion $\Phi$ as the expected per-symbol divergence\footnote{Following for instance \cite[Remark 3]{Chen_RDPAchivability}, one can choose the perception fidelity criterion to be $D( \p{} || q_{\hat{x}_i}), ~i=1,\ldots,n$, or $D( \p{} || \sum_{i=1}^n q_{\hat{x}_i})$. Both types of fidelity criteria result in the same operational quantity for finite alphabets in the asymptotic regime.} as follows:
\begin{align}
\Delta \triangleq \E{\frac{1}{n} \sum_{i=1}^n  d(x_i,\hat{x}_i)}\nonumber \qquad
\Phi \triangleq \frac{1}{n} \sum_{i=1}^n D( \p{} || q_{\hat{x}_i}).\nonumber
\end{align}
We are now ready to introduce the definition of achievability and that of infimum of all achievable rates.
\begin{definition}(Achievability)
Given a distortion level $D > 0$ and a perception constraint $P > 0$, a rate $R$ is said to be $(D,P)$-achievable if for an arbitrary $\epsilon > 0$, there exists, for large enough n, a deterministic lossy source code $(n, M, \Delta, \Phi)$ with $M \le 2^{n(R + \epsilon)}$ such that $\Delta \le D+\epsilon$ and $\Phi \le P+\epsilon$. Then, we define
\begin{align}
R_{nr}(D,P) \equiv \inf\{R: (R,D,P) \text{ is achievable}\}.\label{eq: rate_inferior_nr}
\end{align}
\end{definition}
Next, we give the definition of the information theoretic characterization of the RDPF (see \cite{BlauMichaeli_RDP}) assuming that $D>0$ and $P>0$. 
\begin{definition}(RDPF)\label{def:rdpf}
    For a given finite alphabet source distribution $\p{}$, a single-letter distortion $d(\cdot,\cdot)$ and a divergence $D(\cdot||\cdot)$, the RDPF is characterized as follows:
    %\begin{align} 
        \begin{align}
        \begin{split}
            R(D,P) = & \min_{\Q{}}  I(X,\hat{X}) \\
            \textrm{s.t.}   & \quad \E{d(x,\hat{x})} \le D , \quad D(\p{}||\Qm{}) \le P 
           \end{split}\label{opt: primal}
        \end{align}
    %\end{align}
where $D\in[D_{\min},D_{\max}]\subset(0,\infty)$, $P\in[P_{\min},P_{\max}]\subset(0,\infty)$ and
\begin{align}
    I(X,\hat{X}) = D_{KL}(\p{} \Q{}, \p{} \Qm{}) \triangleq I(\p{}, \Q{}) 
\end{align}
where $I(\p{}, \Q{})$ highlights the dependency on $\{\p{},\Q{}\}$.
\end{definition}
We stress the following technical remarks on Definition \ref{def:rdpf}.
\begin{remark}\label{remark:1}(On Definition \ref{def:rdpf}) Following \cite{BlauMichaeli_RDP}, it can be shown that \eqref{opt: primal} has some useful properties, under mild regularity conditions.
In particular, \cite[Theorem 1]{BlauMichaeli_RDP} shows that $R(D,P)$ is (i) monotonically non-increasing function in both $D\in[D_{\min},D_{\max}]\subset[0,\infty)$ and $P\in[P_{\min},P_{\max}]\subset[0,\infty)$; (ii) convex if the divergence $D(\cdot||\cdot)$ is convex in its second argument. 
\end{remark}
\par In the sequel, we consider that in \eqref{opt: primal}  {\it the perception constraint is an $f$-divergence}, i.e., $D(\cdot||\cdot)=D_f(\cdot||\cdot)$, which is known to be convex in both arguments \cite[Lemma 4.1]{fdiv-csizar}. Hence in light of the discussion in Remark \ref{remark:1}, $R(D,P)$ forms a convex programming problem.
\par We conclude this section by providing a theorem that connects $R_{nr}(D,P)$ with  $R(D,P)$ for finite alphabets, as long as $D>0$ and $P>0$.
\begin{theorem} For $|{\cal X}|<\infty$, $D>0$, $P>0$, we obtain
\begin{align}
R_{nr}(D,P)=R(D,P).
\end{align}    
\end{theorem}
\begin{IEEEproof}
 This is a consequence of \cite[Assumption 2]{Chen_RDPAchivability} that results into \cite[Theorem 2]{Chen_RDPAchivability}.   
\end{IEEEproof}
It should be noted that in the previous analysis, we excluded the extreme case where $P=0$. This is because, in that scenario, deterministic encoders and decoders do not achieve $R(D,0)$ and one instead requires common \cite[Theorem 3]{theis:2021} or private \cite[Theorems 4]{Chen_RDPAchivability} randomness to achieve it. 

\section{Main Results}
In this section, we present our main results. 
We start by reformulating  \eqref{opt: primal} into a double minimization problem. 

\begin{lemma} (Double minimization)\label{lemma: Minimization_Conditions}
Let $s_1\geq{0}$, $s_2\geq{0}$, and define $s=(s_1, s_2)$. Moreover, let $D>0$, $P>0$ and let $D(\cdot||\cdot)=D_f(\cdot||\cdot)$. Then \eqref{opt: primal}  can be expressed as a double minimum as follows:
\begin{align}
\begin{split}
R(D,P) = &-s_1 D - s_2 P+ \min_{\Q{}} \min_{\q{}} \Bigl[ s_1 \E[]{d(x,\hat{x})}\\
& +s_2D_f(\p{}||\Qm{}) + D_{KL}(\p{} \Q{} || \p{} \q{} )\Bigr]
\end{split}\label{eq: double_min}
\end{align}
where  $D =\E[\Q{*}]{d(x,\hat{x})}$, $P=D_f(\p{}||\Qm{*})$, with $\Q{*}$ achieving the minimum. Furthermore, for fixed $\Q{}(\hat{x}|x)$, the right side of \eqref{eq: double_min} is minimized by
\begin{align}
\q{}(\hat{x}) = \sum_{x \in \mathcal{X}}  \p{}(x) \Q{}(\hat{x}|x). \label{eq: optimization qx}
\end{align}
For fixed $\q{}$, the right side of \eqref{eq: double_min} is minimized by
\begin{align}
\Q{}(\hat{x}|x) &= \q{}(\hat{x}) \cdot \tfrac{\A{}}{\sum_{i \in \mathcal{X}} \q{}(i) \A[i]{}}\label{eq: optimization Qxx - Q}
\end{align}
where
\begin{align}
\A{} &= \exp\left\{-s_1 d(x,\hat{x}) - s_2 g(\p{},\Qm{}, \hat{x})  \right\} \label{eq: optimization Qxx - A}\\
g(\p{},\Qm{},\hat{x}) &=  f\left(\tfrac{ \p{}(\hat{x}) }{\Qm{}(\hat{x})}\right) - \tfrac{ \p{}(\hat{x}) }{\Qm{}(\hat{x})} \df\left(\tfrac{ \p{}(\hat{x}) }{\Qm{}(\hat{x})}\right).\nonumber 
\end{align} 
\end{lemma}
%\begin{IEEEproof}
%See Appendix \ref{proof: Minimization Conditions}.
%\end{IEEEproof}
We note that Lemma \ref{lemma: Minimization_Conditions} differs from \cite[Theorem 6.3.3]{Blahut1987PrinciplesAP} in \eqref{eq: optimization Qxx - Q}. Specifically, the presence of the additional perception constraint $D_f(\cdot||\cdot)$ in \eqref{opt: primal} changes the functional properties of the parametric solution \eqref{eq: optimization Qxx - Q}, as it requires an additional exponential term, i.e., $s_2g(\p{},\Qm{}, \hat{x})$, with $g(\cdot)$ depending on the induced marginal $\Qm{}$.
\par The next corollary is a consequence of Lemma \ref{lemma: Minimization_Conditions}.
\begin{corollary} \label{cor: Minimization Conditions}
$R(D,P)$ in \eqref{eq: double_min} can be reformulated as follows:
%\begin{align} 
\begin{align}
\begin{split}
R(D_s,P_s) =& -s_1D_s -s_2P_s\\
&+ \min_{\q{}} s_2 \ssumh{\hat{x}} \p{}(\hat{x}) \df\left(\tfrac{ \p{}(\hat{x}) }{\Qm{}[\q{}](\hat{x})}\right)\\ 
& - \ssum{x} \p{}(x)\log\left(\ssumh{\hat{x}} \q{}(\hat{x}) \A{}\right) 
\end{split}\label{eq: dual after Q optimization}
\end{align}
where $\q{*}$ achieves the minimum and
\begin{align}
    D_s &= \dsum{x}{\hat{x}} \p{}(x) \tfrac{ \q{*}(\hat{x}) \A{}}{\ssumh{i} \q{*}(i) \A[i]{} } d(x,\hat{x})\nonumber\\
    P_s &=  D_f(\p{}||\Qm{}[\q{*}]).\nonumber
\end{align}
\end{corollary}
%\begin{IEEEproof}
%The proof follows by substituting the parametric solution of $\Q{}(\hat{x}|x)$ in \eqref{eq: double_min}.
%\end{IEEEproof}

Next, we proceed to construct an alternating minimization procedure and show its convergence to a point of $R(D,P)$. 
\begin{theorem} (Alternating minimization) \label{th:BA Convergence}
Let $s_1 \ge 0$, $s_2 \ge 0$  be given, with $s=(s_1, s_2)$, and let
$A[\q{}]$ be such that
\begin{align*}
A[\q{}](x,\hat{x},s)&=\exp\left\{-s_1 d(x,\hat{x}) - s_2 g(\p{},\Qm{}[\q{}],\hat{x})\right\}\\
g(\p{},\Qm{}[\q{}],\hat{x}) &=  f\left(\tfrac{ \p{}(\hat{x}) }{\Qm{}[\q{}](\hat{x})}\right) - \tfrac{ \p{}(\hat{x}) }{\Qm{}[\q{}](\hat{x})} \df\left(\tfrac{ \p{}(\hat{x}) }{\Qm{}[\q{}](\hat{x})}\right).\nonumber
\end{align*}
Let $\q{(0)}$ denote any probability vector with nonzero components and let $\q{(n+1)} \equiv\Qm{}[\q{(n)}]$ and $\Q{(n+1)} \equiv\Q{}[\q{(n)}]$ be defined as functions of the previous iteration $\q{(n)}$ as follows:
\begin{align}
&\Q{(n+1)} (\hat{x}|x) = \q{(n)}(\hat{x}) \tfrac{\A{(n)}}{\ssumh{i} \q{(n)}(i) \A[i]{(n)} } \nonumber \\
&\q{(n+1)} (\hat{x}) = \q{(n)}(\hat{x})  \sum_{x \in \mathcal{X}} \tfrac{ \p{}(x) \A{(n)}}{\ssumh{i} \q{(n)}(i) \A[i]{(n)} } 
\label{eq: recursion_q}
\end{align}
where $\A{(n)} = A[\q{(n)}](x,\hat{x},s)$.  Then, as $n\xrightarrow{}\infty$, we obtain
\begin{align*}
D(\Q{(n)})\xrightarrow{} D_s,~P(\Q{(n)})\xrightarrow{} P_s,~I(\p{}, \Q{(n)}) \xrightarrow{} R(D_s,P_s).
\end{align*}
\end{theorem}
%\begin{IEEEproof}
 %   See Appendix \ref{proof: BA Convergence}.
%\end{IEEEproof}
\par Despite being optimal, the alternating minimization scheme of Theorem \ref{th:BA Convergence} does not allow the implementation of a BA type of algorithmic embodiment. Due to the structure of the iterations in \eqref{eq: recursion_q}, a non-reversible dependency of $\q{(n+1)}$ on itself appears once $\A{(n)}$ is substituted therein, thus requiring the knowledge of $\q{(n+1)}$ to evaluate $\q{(n+1)}$ itself. To circumvent this technical difficulty, we introduce a relaxation in the structure of the iterations of Theorem \ref{th:BA Convergence}.  
This results in a variant of an approximate alternating minimization scheme, for which we derive necessary and sufficient conditions that ensure its convergence to a globally optimal point. The aforementioned relaxation and the necessary and sufficient conditions that lead to a globally optimal solution are stated in the following theorem, which is a major result of this paper.
\begin{theorem}(Approximate alternating minimization) \label{th:BA Convergence Approximate}    Let $s_1 \ge 0$, $s_2 \in [0, s_{2, \max}]$ be given with $s=(s_1,s_2)$ and let  $\tilde{A}[\q{}](x,\hat{x},s)$ be
\begin{align}
\tilde{A}[\q{}](x,\hat{x},s)=\exp\{-s_1 d(x,\hat{x}) - s_2 \tilde{g}(\p{},\v{}[\q{}],\hat{x})\} \nonumber \\
\tilde{g}(\p{},\v{}[\q{}],\hat{x}) =  f(\tfrac{ \p{}(\hat{x}) }{\v{}[\q{}](\hat{x})}) -\tfrac{ \p{}(\hat{x})} {\v{}[\q{}](\hat{x})} \df(\tfrac{ \p{}(\hat{x}) }{\v{}[\q{}](\hat{x})}) \nonumber
\end{align}
where $\v{}[\q{}]$ is any probability vector. Let $\tq{(0)}$ be any probability vector with nonzero components and let $\tq{(n+1)} = \Qm{}[\tq{(n)}]$ and $\tQ{(n+1)} = \Q{}[\tq{(n)}]$ be defined as functions of the past iteration $\tq{(n)}$ as follows:
\begin{align*}
    \tQ{(n+1)}  & = \tq{(n)}(\hat{x}) \tfrac{\tA{(n)}}{ \ssumh{i} \q{(n)}(i) \tA[i]{(n)}}\\
    \tq{(n+1)}  & = \tq{(n)}(\hat{x})  \sum_{x \in \mathcal{X}} \tfrac{ \p{}(x) \tA{(n)}}{\ssumh{i} \tq{(n)}(i) \tA[i]{(n)}} 
\end{align*}
where $\tA{(n)} = \tilde{A}[\tq{(n)}](x,\hat{x},s)$. Then, as $n\xrightarrow{}\infty$, we obtain
\begin{align*}
D(\tQ{(n)}) \xrightarrow{} D_s,~P(\tQ{(n)}) &\xrightarrow{} P_s,~I(\p{}, \tQ{(n)}) \xrightarrow{} R(D_s,P_s)
\end{align*}
if and only if $\displaystyle \lim_{n\rightarrow\infty}\tfrac{\tq{(n+1)}}{\v{}[\tq{(n)}]} \to 1$ with at least linear rate of convergence.
\end{theorem}
%\begin{IEEEproof}
%See Appendix \ref{proof:BA Convergence Approximate}.
%\end{IEEEproof}
Theorem \ref{th:BA Convergence Approximate} enables the implementation of the alternating minimization scheme by introducing an auxiliary variable $\v{}[\q{(n)}]$ (approximation), designed as a function of only the current iteration of $\q{(n)}$. Nevertheless, depending on $\v{}$, this approximation may incur restrictions on the domain of the Lagrangian multiplier $s_2$ in order to have convergence guarantees. The implementation of Theorem \ref{th:BA Convergence Approximate} is illustrated in Algorithm \ref{alg: generic SBA}.
\par We conclude this section, with a lemma where we provide necessary and sufficient conditions for the optimal solution of the studied problem.
\begin{lemma} \label{lemma: opt_condition_q}
Let $D_f(\cdot || \cdot)$ be such that $f \in C^1(0, \infty)$ continuous and differentiable on $(0, \infty)$. Then, a probability vector $\q{}$ yields a point on the $R(D,P)$ curve via the transition matrix
\begin{align*}
    \Q{}[\q{}](\hat{x})= \q{}(\hat{x}) \tfrac{ \p{}(x) \tA{}}{\ssumh{i} \q{}(i) \tA[i]{} }
\end{align*}
if and only if, $\forall \hat{x} \in \mathcal{\hat{X}}$, 
\begin{align*}
    \cx[] = \ssum{x} \tfrac{ \p{}(x) \tA{}}{\ssumh{i} \tq{}(i) \tA[i]{}} \le 1,
\end{align*} 
holding with equality for any $\hat{x}$ for which $\q{}(\hat{x})$ is nonzero.
\end{lemma}
%\begin{IEEEproof}
 %See Appendix \ref{proof: opt_condition_q}.
%\end{IEEEproof}

\section{Analysis of Algorithm \ref{alg: generic SBA}}

\par In this section, we study a stopping criterion and the convergence rate of Algorithm \ref{alg: generic SBA}.

\begin{algorithm}
    \caption{Implementation of Theorem \ref{th:BA Convergence Approximate}} \label{alg: generic SBA}
    \begin{algorithmic}[1]
        
        \Require source distribution $\p{}$; Lagrangian multipliers $s = (s_1, s_2)$ with $s_1 \ge 0$ and $s_2 \in [0, s_{2,\max}]$; error tolerance $\epsilon$; divergence measure $D_f(\cdot||\cdot)$; distortion measure $d(\cdot,\cdot)$; initial assignment $\q{(0)}$.
        
        \State $n \gets 0$
        \State $\flag \gets 0$
        
        \While{$\flag == 0$}
            \State  $g(\hat{x}) \gets f \left(\tfrac{\p{}(\hat{x}) }
            {\v{(n)}(\hat{x})} \right) - \tfrac{\p{}(\hat{x}) }{\v{(n)}(\hat{x})} \df \left( \tfrac{\p{}(\hat{x}) }{\v{(n)}(\hat{x})} \right)$ 
            \State $\tA{(n)} \gets \exp \left[-s_1 d(x,\hat{x}) + s_2 g(\p{}, \v{}[\q{}], \hat{x}) \right]$
            \State $c^{(n)}(\hat{x}) \gets \ssum{x} \tfrac{ \p{}(x) \tA{(n)}}{\ssum{i} \tq{(n)}(i) \tA[i]{(n)}}$
            \State $\q{(n+1)} \gets \q{(n)} \cdot c^{(n)}$
            \State $\omega \gets  \log c^{(n)}_{\max}(\hat{x}) - \ssumh{\hat{x}} \q{(n)} c^{(n)}(\hat{x}) \log(c^{(n)}(\hat{x}))$
        \If{$\omega \le \epsilon$}
            \State $\flag \gets 1$
        \EndIf
        
        \State $n \gets n + 1$

        \EndWhile
        \Ensure 
        { $\Q{} = \q{(n)}(\hat{x}) \tfrac{\tA{(n)}}{\ssumh{i} \q{(n)}(i) \tA[i]{(n)} }$},
        { $D_s = \mathbf{E}_{\p{} \Q{}}[d(x,\hat{x})]$},
        { $ P_s = D_f(\p{},\q{(n)}) $},
        {  $R(D_{s},P_s) =  \tW[\q{(n)}]-s_1D_s -s_2P_s  - \ssumh{\hat{x}} \q{(n)} \cx \log(\cx[])$, $\tW{}[\cdot]=\eqref{eq: W_tilde}$.}

    \end{algorithmic}
\end{algorithm}

Before obtaining stopping conditions for Algorithm \ref{alg: generic SBA}, we first derive a useful lemma which is needed to obtain the stopping conditions. 
\begin{lemma} \label{lemma: RDP lowerbound}
 Let $s_1 \ge 0$, $s_2 \in [0, s_{2, \max})$ be given with $s=(s_1,s_2)$ and let $\Q{}$ be a transition matrix included in the set $\mathcal{L}_{(D,P)}$ defined as follows:
\begin{align*} 
    \mathcal{L}_{(D,P)} = \{ \Q{}: \E[\Q{}]{d(x,\hat{x})} \le D \land D_f(\p{}|| \Qm{}) \le P \}.
\end{align*}
Then, $\forall \mathbf{\lambda} \in \Lambda_{s,\v{}[\Qm{}]}$, with $\Lambda_{s,\v{}[\Qm{}]}= \big\{ \lambda \in \mathcal{R}^{|\mathcal{X}|}: \forall x \in \setX, \lambda(x) \ge 0  \land \forall \hat{x} \in \setXh,~\ssum{x} \p{}(x) \lambda(x) \tA{} \le 1 \big\}$, we obtain
\begin{align*} 
\begin{split}
    R(D,P) \ge & - \ssum{x} \p{}(x) \log\left(\tfrac{1}{\lambda(x)}\right) - s_1 D \\
                & - s_2 \dsum{x}{\hat{x}} \p{}(x) \Q{}(x,\hat{x}) \tilde{g}(\hat{x}).
\end{split}
\end{align*}
\end{lemma}
%\begin{IEEEproof}
 %   See Appendix \ref{proof: RDP lowebound}.
%\end{IEEEproof}
\begin{theorem} (Stopping criterion) \label{th: Stopping Conditions}
Let $\tQ{}$ and $\tq{}$ be defined  as in Theorem \ref{th:BA Convergence Approximate}, $\cx[]$ be as defined as in Lemma \ref{lemma: opt_condition_q} and $c_{\max} = \max_{\hat{x} \in \setXh} \cx[]$. Then, at the point $D=\E[\tQ{}]{d(x,\hat{x})}$, and $P=D_f(\p{}||\tQm{}))$, the following bounds hold
\begin{align} 
    R(D,P) &\ge -s_1D -s_2P + \tW[\tq{}] - \log(c_{\max}) \label{eq: Lowerbound Approximation}\\
    \begin{split}
        R(D,P)  & \le -s_1D -s_2P + \tW[\tq{}]  \\
                & \quad - \ssumh{\hat{x}} \tq{} \cx \log(\cx[]) \label{eq: Upperbound Approximation}
    \end{split}
\end{align}

where $\tW{}[\tq{}]$ is given by 
\begin{align}
   \begin{split}
    \tW[\tq{(n)}] &= - \ssum{x} \p{}(x) \log \left( \ssumh{\hat{x}} \tq{(n)}(\hat{x})\tA{(n)} \right)\\
    & \quad +s_2 \ssumh{\hat{x}} \p{}(\hat{x}) \tfrac{ \p{}(\hat{x}) }{\v{(n)}(\hat{x})} \df \left(\tfrac{ \p{}(\hat{x}) }{\v{(n)}(\hat{x})} \right)\\
    & \quad + s_2 \Bigg{[} \ssumh{\hat{x}} \tq{(n+1)} \left(f \left(\tfrac{ \p{}(\hat{x}) }{\q{(n+1)}(\hat{x})} \right) - f \left(\tfrac{ \p{}(\hat{x}) }{\v{(n)}(\hat{x})} \right) \right) \Bigg{]}.
    \end{split} \label{eq: W_tilde}
\end{align}

\end{theorem}
%\begin{IEEEproof}
 %   See Appendix \ref{proof: Stopping Conditions}.
%\end{IEEEproof}
\par Next, we study the convergence rate of Algorithm \ref{alg: generic SBA}. This is done by first studying the convergence rate of the alternating minimization procedure of Theorem \ref{th:BA Convergence} and then, using it as a reference to analyze the convergence rate of the approximate alternating minimization procedure of Theorem \ref{th:BA Convergence Approximate}. 
\par Note that based on the structure of the parametric solution of $\q{(n+1)}$ at the previous iteration $n$, we can similarly define a vector function $S: \mathcal{R}^{|\mathcal{X}|} \to \mathcal{R}^{|\mathcal{X}|}$ as $S[\q{}](i) = \q{}(i) \cdot c[\q{}](i)$. Using Lemma \ref{lemma: opt_condition_q}, a distribution $\q{*}$ that achieves the RDPF is a fixed point of $S(\q{})$. Following \cite{9476038}, we can analyze the convergence rate of the alternating minimization procedure in both Theorems \ref{th:BA Convergence} and \ref{th:BA Convergence Approximate}. Using the first order Taylor expansion of $S[\q{}]$ around a fixed point $\q{*}$, we obtain
%\begin{align*}
 $ S[\q{}] = S[\q{*}] + J(\q{*}) \cdot (\q{} - \q{*}) + o(||\q{} - \q{*}||)$,
%\end{align*}
where $J(\q{})$ is the Jacobian matrix of $S(\q{})$ with entries $
    J_{i,j}(\q{}) \triangleq \frac{\partial S[\q{}](i)}{\partial \q{}(j)}, (i,j) \in \setX \times \setXh.$
The next theorem provides the functional form of the Jacobian for the case of Theorem \ref{th:BA Convergence}.
\begin{theorem}(Jacobian form) \label{th: base_alg_Jacobian}
The Jacobian $J(\q{})$ computed at the fixed point $\q{*}$ is given as 
    \begin{align} 
        J(\q{*}) = (I - M)(I - \Gamma J(\q{*})) \label{eq: JacobianEq1}
    \end{align}
where
 \begin{align}
     M &\triangleq \left[ \q{*}(i) \ssum{x} \p{}(x) \tfrac{\A[i]{} \A[j]{}}{(\ssumh{k} \q{*}(k) \A[k]{})^2}  \right]_{(i,j) \in \setX \times \setXh} \label{matrix: M}\\
     \Gamma &\triangleq s_2 \cdot \diag\left[ \q{*}(i) \cdot \tfrac{\partial^2}{\partial q(i)^2} D_f(\p{}||q) \bigg{|}_{\q{*}} \right]_{i \in \mathcal{X}}. \label{matrix: Gamma}
\end{align} 
\end{theorem}
%\begin{IEEEproof}
 %   See Appendix \ref{proof: base_alg_Jacobian}.
%\end{IEEEproof}
Next, we introduce two lemmas, in which we use the structure of \eqref{eq: JacobianEq1} to identify properties of matrix $M$.  

\begin{lemma} \label{lemma: lowerboundEigM}
Let $\{ \lambda_i \}_{i = 1:|\setX|}$ be the set of eigenvalues of $M$. Then, given a distortion function $d: \mathcal{X} \times \mathcal{ \hat{X} } \to \mathcal{R}_0^+$ that induces a full-rank matrix $D=[e^{ - s_1 d(i,j)}]_{(i,j) \in \setX \times \setXh}$, then $\lambda_i > 0$,$\forall i \in [1:|\mathcal{X}|]$, i.e., M has only positive eigenvalues.  
\end{lemma}
%\begin{IEEEproof}
 %   See Appendix \ref{proof: loweboundEigM}.
%\end{IEEEproof}

\begin{remark}(On Lemma \ref{lemma: lowerboundEigM}) We note that a popular example that satisfies the assumptions imposed on Lemma \ref{lemma: lowerboundEigM} is the Hamming distortion denoted hereinafter as $d_H$ \cite{cover-thomas:2006}.
\end{remark}

\begin{lemma} \label{lemma: upperboundEigM}
Let $\{ \lambda_i \}_{i = 1:|\setX|}$ be the set of eigenvalues of $M$. Then, at fixed point $\q{*}$, we have that $\lambda_i \le 1, \forall i \in 1:|\setX|.$
\end{lemma}
%\begin{IEEEproof}
 %   See Appendix \ref{proof: upperboundEigM}
%\end{IEEEproof}

Using Lemmas \ref{lemma: lowerboundEigM} and \ref{lemma: upperboundEigM}, we can now characterize the interval that contains the set of eigenvalues of $J(\q{*})$ and subsequently the convergence rate of Theorem \ref{th:BA Convergence}. 
\begin{theorem}(Convergence rate of Theorem \ref{th:BA Convergence}) \label{th: EXPConvergenceBase}
    Let $\{\theta_i \}_{i \in \mathcal{X}}$ be the eigenvalues of $J(\q{*})$. Then,
    \begin{align*}
        0  \le \{\theta_i \}_{i \in \mathcal{X}} < 1.
    \end{align*}
    Moreover, let $\gamma \in [\theta_{\max}, 1)$. Then, there exists $\delta > 0$ and $K > 0$ such that if $\q{0} \in \{\q{}: ||\q{} - \q{*}|| \le \delta \}$, we obtain
    \begin{align}
        ||\q{(n)} -  \q{*}|| < K \cdot || \q{(0)} - \q{*}|| \cdot \gamma^n
    \end{align}
    i.e., the iterations converge exponentially.
\end{theorem}
%\begin{IEEEproof}
% See Appendix \ref{proof: EXPConvergenceBase}.
%\end{IEEEproof}

Following similar steps that led to Theorem \ref{th: EXPConvergenceBase}, we can now write the Jacobian $J_a(\q{*})$ for Theorem \ref{th:BA Convergence Approximate} considering a specific form of the auxiliary variable $\v{}[\q{(n)}] = \q{(n)}$. This results into the following structure of the Jacobian matrix
\begin{align*}
    J_a(\q{*}) = (I - M)(I - \Gamma)
\end{align*}
where $M$ and $\Gamma$ are given by \eqref{matrix: M} and \eqref{matrix: Gamma}, respectively. Unlike Theorem \ref{th:BA Convergence}, where the structure of \eqref{eq: JacobianEq1} bounds its own eigenvalues, in this case we need to bound the Lagrangian multiplier $s_2$, hence matrix $\Gamma$, to guarantee exponential convergence of the algorithm. This is proved in the following theorem.
\begin{theorem} \label{th: approximation_exponential_interval}
    For a given $s_1\ge{0}$, let $I_{s_2} = [0, s_{2,\max}]$ be the domain of $s_2$, $\{\theta_{a,i}\}_{i \in \mathcal{X}}$ the set of eigenvalues of $J_a(\q{*})$ and $\theta_{\max}$ the maximum eigenvalue of $J(\q{*})$ in \eqref{eq: JacobianEq1}. Define the set $I_{s_2}^\epsilon = [0, s_{2,\max} - \epsilon]$ for $0 < \epsilon < s_{2,\max} $. Then, there exists an $\epsilon^\prime$ such that if $s_2 \in I_{s_2}^{\epsilon^\prime}$ then $0 \le \{\theta_{a,i}\}_{i \in \mathcal{X}} < 1$. 
\end{theorem}
%\begin{IEEEproof}
 %   See Appendix \ref{proof: approximation_exponential_interval}.
%\end{IEEEproof}
Theorem \ref{th: approximation_exponential_interval} guarantees exponential convergence for Theorem \ref{th:BA Convergence Approximate} only for $s_2 \in  I_{s_2}^{\epsilon}$ which means that we are able to consider $P \in [P_{\min}(\epsilon), P_{\max}]$, depending on the characteristics of the specific problem. 

\section{Numerical Results} \label{sec: Numerical Examples}
In this section, we provide numerical results to demonstrate the utility of Algorithm \ref{alg: generic SBA}.

\begin{figure}[H]
	\centering
    \begin{subfigure}{\linewidth}
        \centering \includegraphics[width=0.4\linewidth]{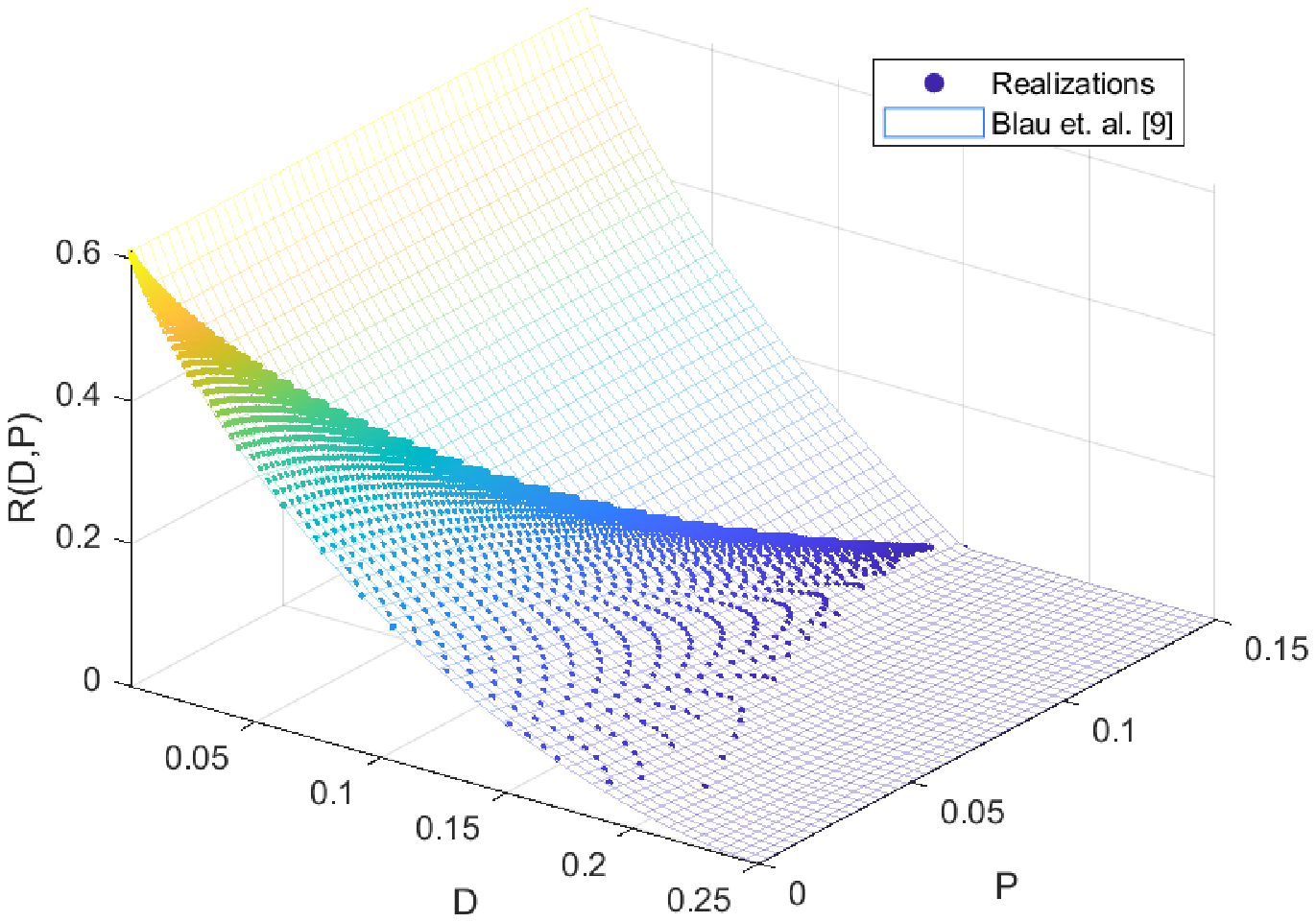}
        \caption{} \label{fig: Bern_TV}
    \end{subfigure}\\[1ex]
	\begin{subfigure}{0.4\linewidth}
		\centering \includegraphics[width=\linewidth]{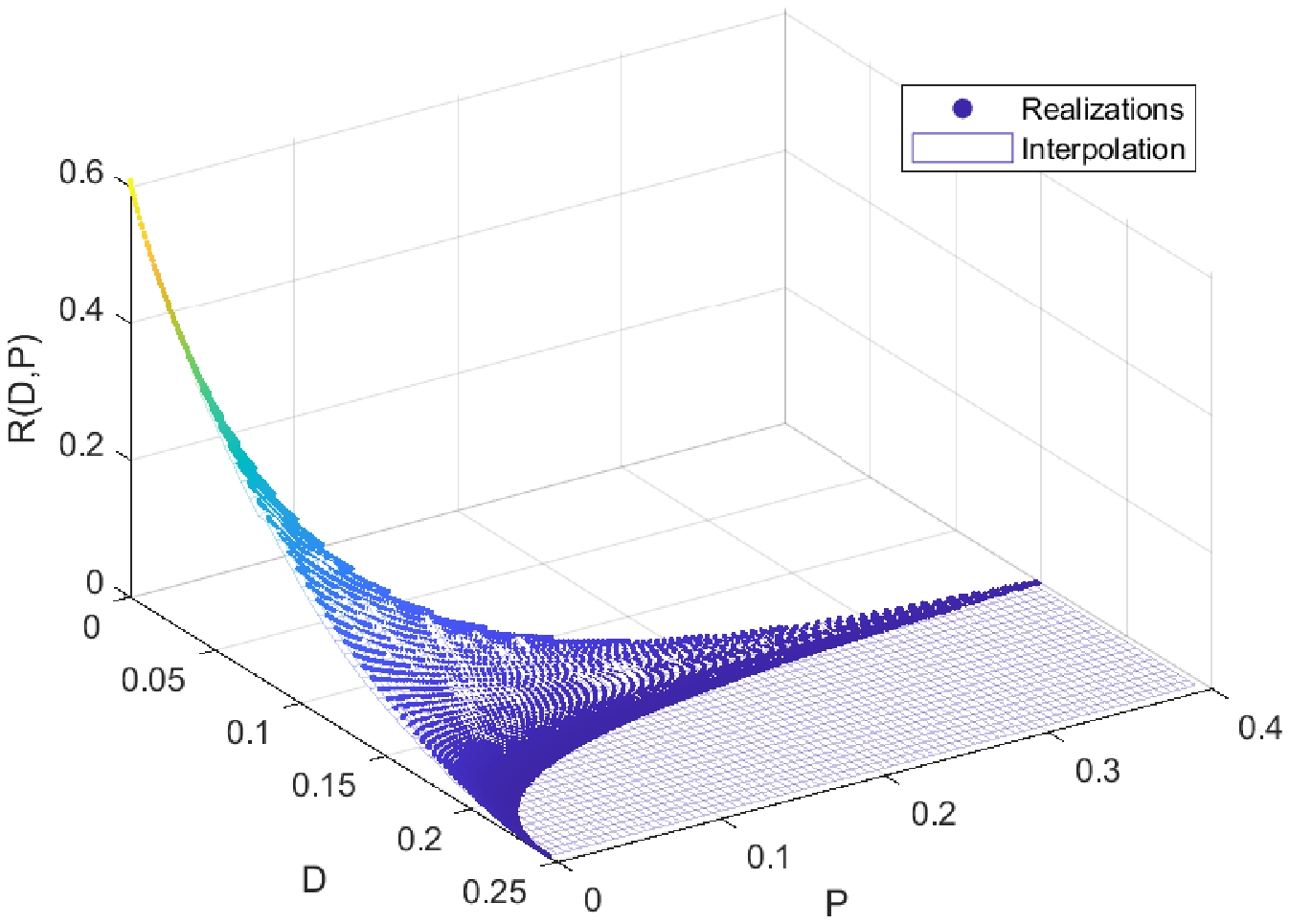}
		\caption{} \label{fig: Bern_KL}
	\end{subfigure}
   	\begin{subfigure}{0.4\linewidth}
     	\centering \includegraphics[width=\linewidth]{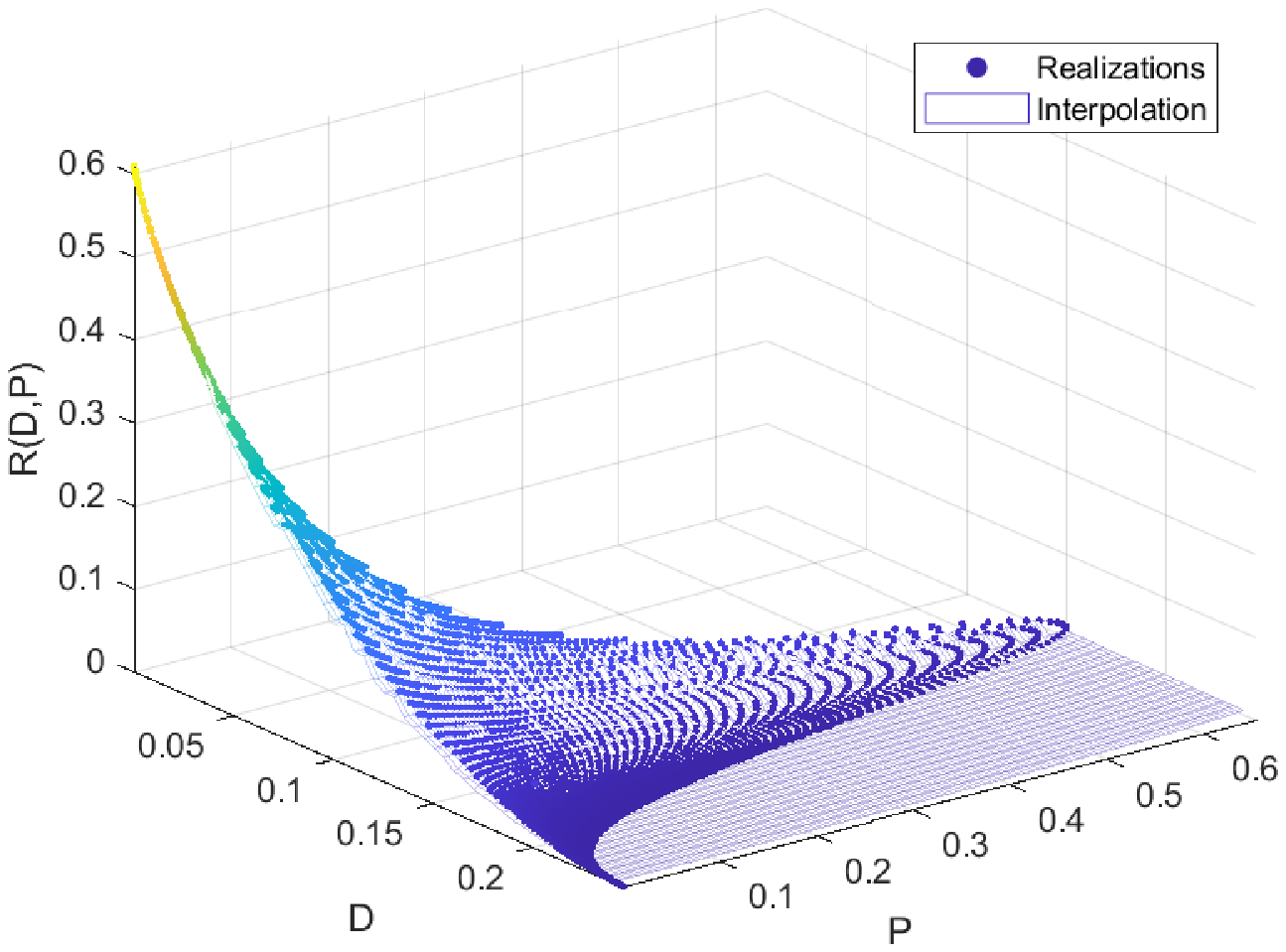}
     	\caption{} \label{fig: Bern_CHI}
   	\end{subfigure}
\caption{$R(D,P)$ for a Bernoulli source under a Hamming distortion and various perception constraints.}
\label{fig:test}
\end{figure}

\begin{example}
Suppose that ${\cal X}=\hat{\cal X}=\{0,1\}$ and let $\p{} \sim Ber(0.15)$ with $d(\cdot,\cdot)=d_H(\cdot,\cdot)$ and perception constraint chosen to be either (a) $ D_f(\p{}||\q{}) = TV(\p{}||\q{}) = \frac{1}{2} \ssum{i} |\p{}(i)-\q{}(i)|$, or (b) $D_f(\cdot||\cdot)=D_{KL}(\cdot||\cdot)$, or (c) $D_f(\cdot||\cdot)=D_{\chi^2}(\cdot||\cdot)$. 
\par In Fig. \ref{fig: Bern_TV} we compare the theoretical results of \cite[Equation 6]{BlauMichaeli_RDP} with the numerical results obtained using Algorithm \ref{alg: generic SBA}. We observe that Algorithm \ref{alg: generic SBA} achieves exactly the theoretical solution of \cite[Equation 6]{BlauMichaeli_RDP} as long as $D\le{D}_{\max}=0.15$. 
\par In Fig. \ref{fig: Bern_KL} and Figure \ref{fig: Bern_CHI} we use Algorithm \ref{alg: generic SBA} to compute $R(D,P)$ for $D_f(\cdot||\cdot)=D_{KL}(\cdot||\cdot)$, and for $D_f(\cdot||\cdot)=D_{\chi^2}(\cdot||\cdot)$, respectively.
\end{example}
\section*{Acknowledgement}\label{ack}
This work has received funding from the European Research Council (ERC) under the European Union’s Horizon 2020 Research and Innovation programme (Grant agreement No. 101003431).

%\clearpage

\IEEEtriggeratref{13}  %% Use it for polishing purproses

\bibliographystyle{IEEEtran}
\bibliography{string,biblio}

\clearpage

\end{document}